\documentclass[%
 reprint, amsmath,amssymb,nofootinbib,
 aps]{revtex4-2}

\usepackage{graphicx}
\usepackage{dcolumn}
\usepackage{bm}
\usepackage{comment}
\usepackage{caption}
\usepackage{subcaption}
\usepackage{float}
\usepackage[usenames,dvipsnames]{xcolor}

\begin{document}

\preprint{APS/123-QED}

\title{Solving the initial conditions problem for modified gravity theories}

\author{Sam E. Brady}
\author{Llibert Aresté Saló}
\author{Katy Clough}
\author{Pau Figueras}
\affiliation{Centre for Geometry, Analysis and Gravitation, School of Mathematical Sciences, Queen Mary University of London,
Mile End Road, London E1 4NS, United Kingdom}
\author{Annamalai P. S.}
\affiliation{CP3-Origins, University of Southern Denmark, Odense, Denmark}

\date{\today}

\begin{abstract}
Modified gravity theories such as Einstein scalar Gauss Bonnet (EsGB) contain higher derivative terms in the spacetime curvature in their action, which results in modifications to the Hamiltonian and momentum constraints of the theory. In principle, such modifications may affect the principal part of the operator in the resulting elliptic equations, and so further complicate the already highly non-linear, coupled constraints that apply to the initial data in numerical relativity simulations of curved spacetimes. However, since these are effective field theories, we expect the additional curvature terms to be small, which motivates treating them simply as an additional source in the constraints, and iterating to find a solution to the full problem. In this work we implement and test a modification to the CTT/CTTK methods of solving the constraints for the case of the most general four derivative, parity invariant scalar-tensor theory, and show that solutions can be found in both asymptotically flat/black hole and periodic/cosmological spacetimes, even up to couplings of order unity in the theory. Such methods will allow for numerical investigations of a much broader class of initial data than has previously been possible in these theories, and should be straightforward to extend to similar models in the Horndeski class.

\end{abstract}

\maketitle

\section{\label{sec:level1}Introduction}

Recent breakthroughs in well posed formulations \cite{Reall2020_1, Reall2020_2} have resulted in an expansion of the class of modified theories of gravity to which numerical relativity (NR) -- numerical simulations that solve the Einstein Equations as a time evolution problem -- can be applied. This has allowed for the simulation of strong gravity spacetimes in these theories, including the fully non-linear backreaction of the additional curvature terms onto the metric \cite{East:2020hgw,East:2021bqk,East:2022rqi,Corman:2022xqg,AresteSalo2022,AresteSalo23}, building on previous works that neglected such effects \cite{Richards:2023xsr,R:2022tqa,Okounkova:2022grv,Elley:2022ept,Doneva:2022byd,Okounkova:2020rqw,Silva:2020omi,Okounkova:2019zjf,Okounkova:2019dfo,Witek:2018dmd}. Such simulations are of particular interest for the construction of binary merger waveforms, but also have relevance to more general questions about the hyperbolicity of such theories in the strongly coupled regime \cite{East:2020hgw,East:2021bqk,East:2022rqi,R:2022hlf, Doneva23,Thaalba:2023fmq}, and potentially in early Universe cosmology \cite{Nojiri:2017ncd}. However, in works to date in 3+1 spacetime dimensions, the challenge of satisfying the more complicated constraint equations has limited the physical scenarios that can be investigated. Most have begun with a trivial solution that satisfies the regular constraint equations of general relativity (GR) and set the additional scalar degree of freedom to zero (it is then evolved to a non trivial configuration over time) \cite{East:2020hgw,East:2022rqi,Corman:2022xqg,AresteSalo2022}. Alternatively, studies of spin-induced scalarisation that require a scalar ``seed'' have tolerated a certain level of constraint violation in the initial data arising from a non trivial profile of the scalar, and relied on constraint damping to remove it during the initial stages of the evolution \cite{East:2021bqk,Doneva23,AresteSalo23}.

The Arnowitt-Deser-Misner (ADM) \cite{ADM59} decomposition of the Einstein Field Equations provides the basis for NR simulations, posing the system as a Cauchy problem, with a set of evolution and constraint equations for the 3-dimensional spatial metric $\gamma_{ij}$ and the extrinsic curvature, $K_{ij}$, of the spatial slices.
The ADM approach gives rise to four independent constraint equations, which must be satisfied on each time slice, and in particular in the initial data that starts an NR simulation. These constraints are given by
\begin{align}
{\mathcal H}\equiv R+K^2-K_{ij}K^{ij} - 16\pi\rho=0 \label{eq:Ham} ~,\\
{\mathcal M}_i\equiv D_jK_{~i}^j-D_iK - 8\pi S_i=0 \label{eq:Mom} ~,
\end{align}
where $K=\gamma^{ij}K_{ij}$ is the trace of the extrinsic curvature (also known as the mean curvature), and $R$ and $D_i$ are the Ricci scalar and covariant derivative associated with the 3-dimensional spatial metric. The energy density $\rho$ and momentum densities $S_i$ are functions of the matter fields and their derivatives. These can be derived directly from the matter action as projections of the stress-energy tensor $T_{\mu\nu}$. In the case of modified gravity theories like EsGB, the additional curvature terms can be treated as further effective matter sources that depend on higher order derivatives of the other metric variables.

Including the lapse $\alpha$ and shift $\beta^i$, there are 16 components that must be specified initially, but only four constraint equations. Whilst four of the extra components relate to physical degrees of freedom, the remainder are gauge. Values that represent free data and those that are fully determined by the physical scenario are not easy to separate, and so some must be chosen somewhat arbitrarily. The result is that there will often be a large number of possible ways to set the free data that appear to still meet the physical requirements. However, poor choices can result in uniqueness and existence problems in the initial constraints that prevent unique solutions being found.

Many different approaches have been developed for solving the constraint equations (for reviews see the standard NR texts \cite{Alcubierre08, Baumgarte:2010ndz, 2016nure.book.....S}). These approaches vary in the degrees of freedom that have their values imposed, and those that are solved for. The metric and extrinsic curvature can be decomposed in various ways (e.g., through a conformal decomposition), allowing for the freely chosen variables to be more closely identified with a particular physical interpretation. This can also leave the equations in a form more amenable to numerical solutions.

One such approach is known as the Conformal-Transverse-Traceless (CTT) method. In this approach the extrinsic curvature is decomposed into its trace $K$ and a traceless tensor $A_{ij}$. The 3-metric $\gamma_{ij}$ is decomposed into a conformal metric with determinant one $\bar\gamma_{ij}$, and a conformal factor $\psi$, as $\gamma_{ij}=\psi^4\bar\gamma_{ij}$. $A_{ij}$ is similarly decomposed as $A_{ij} = \psi^{-2}\bar A_{ij}$, and the conformal $\bar A_{ij}$ is further decomposed into a transverse-traceless part $\bar{A}^{TT}_{ij}$ and a vector potential $W_i$ (see equation \ref{eq:Aij_decomp}). The conformal metric $\bar\gamma_{ij}$, the mean curvature $K$, and the transverse-traceless  extrinsic curvature $\bar{A}^{TT}_{ij}$ are then imposed to have some values (usually trivially those of a flat spacetime), and the constraints are used to solve for the conformal factor $\psi$ and the vector potential $W_i$. $\bar A_{ij}$ can then be reconstructed as,
\begin{equation}\label{eq:Aij_decomp}
    \bar A_{ij} = \bar A^{TT}_{ij} + \bar D_i W_j + \bar D_j W_i - \frac{2}{3}\bar\gamma_{ij}\bar D_k W^k ~.
\end{equation}
In this method, the term in the energy density $\rho$ is generically problematic, due to the nature of the resulting elliptic equation for $\psi$. It is normally only possible to specify a conformally-related density $\bar\rho$, as this allows for the term in $\rho$ to appear with the ``right sign'' to guarantee unique solutions of the elliptic equation for $\psi$ \cite{Baumgarte06}.

A modification to the CTT technique, known as CTTK, was recently proposed to address this limitation, which is more acute for problems involving fundamental fields as opposed to fluids \cite{Aurrekoetxea22}. In this approach the variables are decomposed in the same way, but the elliptic equation for $\psi$ is split into an algebraic equation for $K$ in terms of $\rho$, and $\psi$ is then only required to satisfy Laplace's equation with a source term in $\bar{A}_{ij}$. The conformal metric $\bar\gamma_{ij}$ and $\bar{A}^{TT}_{ij}$ still need to be chosen, but the mean curvature $K$ is now solved for as part of the algorithm and this simplifies the solution for $\psi$. Overall the method was found to be highly robust, with unique solutions being found by the solver over a wide range of scenarios.

In this work, the CTTK method has been modified to solve the full constraint equations of the most general four derivative, parity invariant scalar-tensor theory of gravity ($4\partial ST$). It has been shown that the CTT formulation of the elliptic equations for these theories has unique solutions in the weak coupling limit \cite{Kovacs:2021lgk}, and one can expect this to carry over to the CTTK case. To achieve this, as suggested in \cite{Kovacs:2021lgk}, the contributions from the higher-derivative curvature terms in the action are treated as additional sources that are split between the available degrees of freedom. The CTTK technique is modified in two different ways, depending on the boundary conditions of the problem at hand, to ensure that solutions exist. We describe the methods in more detail below, and demonstrate their effectiveness in practise by showing that the solutions converge as expected.

We follow the conventions in Wald's book \cite{Wald:1984rg}. Greek letters $\mu,\nu\,\ldots$ denote spacetime indices and they run from 0 to 3; Latin letters $i,j,\ldots$ denote indices on the spatial hypersurfaces and they run from 1 to 3. We set $G=c=1$. 

\section{Methods}
\subsection{Additional terms}
In $4\partial ST$ gravity, the Einstein-Hilbert action is modified to include a scalar field coupled non-trivially to gravity, through the Gauss-Bonnet term $\mathcal{L}^{\text{GB}}$,
\begin{equation}\label{eq:action}
S_{4\partial ST} = \int d^4x \sqrt{-g}\big[ \frac{R}{16\pi} -V(\phi) + X + g_2(\phi)X^2 + \lambda(\phi)\mathcal{L}^{\text{GB}}\big] ~,
\end{equation}
where $X = - \frac{1}{2}(\nabla_\mu\phi)(\nabla^\mu \phi)$,
\begin{equation}
    \mathcal{L}^{\text{GB}} = R^2-4R_{\mu\nu}R^{\mu\nu}+R_{\mu\nu\rho\sigma}R^{\mu\nu\rho\sigma} ~,
\end{equation}
and $g_2(\phi)$ and $\lambda(\phi)$ are smooth functions of $\phi$. This is the most general parity-invariant scalar-tensor theory of gravity that includes up to four derivatives of the metric, and is an example of the wider class of Horndeski theories \cite{Horndeski1974}. The local magnitude of the coupling $\lambda(\phi)$ controls the deviation of the solutions from the GR case, with the $g_2(\phi)$ term being generally subdominant for similar coupling constants and the same field amplitude. Deviations are also amplified in regions of high curvature (i.e., with a larger Gauss-Bonnet invariant).

The effects of these additional terms can be included in $\rho$ and $S_i$, so that the constraints remain as they are in Equations \ref{eq:Ham} and \ref{eq:Mom} but with an effective $\rho$ and $S_i$ given by
\begin{eqnarray}\label{eq:rho_Si_new}
\rho&=&\rho^{\text{SF}}+\rho^X+\rho^{\text{GB}}~, \\
S_i&=&S_i^{\text{SF}}+S_i^X+S_i^{\text{GB}} ~.
\end{eqnarray}
Here $\rho^{\text{GB}}$ and $S_i^{\text{GB}}$ include higher derivative contributions from the scalar and tensor metric variables, and $\rho^X$ and $J_i^X$ contain the contributions of the $g_2(\phi)$ term. Explicitly:
\begin{subequations}\label{edgbcomp}
\begin{eqnarray}
\rho^{\text{GB}}&=&\Omega M - 2M_{kl}\Omega^{kl}\,, \\
S^{\text{GB}}_i&=&\Omega_iM-2M_{ij}\Omega^j\,, \nonumber\\
&&- 4\big(\Omega^j_ {~[i}N_{j]}-\Omega^{jk}D_{[i}K_{j]k}\big)\,, \\
\rho^{\text{X}}&=&\frac{g_2(\phi)}{4}\left( \Pi^2 - D_i\phi D^i\phi\right)\left( 3\Pi^2 + D_i\phi D^i\phi\right), \\
S^{\text{X}}_i&=&-g_2(\phi)\,\Pi\, D_i\phi\left( \Pi^2 - D_i\phi D^i\phi\right)\,,
\end{eqnarray}
\end{subequations}
with
\begin{subequations}
\begin{eqnarray}\label{MNeq}
\hspace{-0.5cm}M_{ij}&=&R_{ij}+\tfrac{2}{9}\gamma_{ij}K^2+\tfrac{1}{3}KA_{ij}-A_{ik}A^k_j\,, \\
\hspace{-0.5cm}N_i&=&-\frac{2}{3}D_iK + D_jA_i^j\,, \\
\hspace{-0.5cm}\Omega_i&=&-4\lambda'\big(D_i\Pi+A^j_{~i}D_j\phi+\tfrac{K}{3}D_i\phi \big)-4\lambda''\Pi\,D_i\phi,\\
\hspace{-0.5cm}\Omega_{ij}&=&4\lambda'\left(D_iD_j\phi+\Pi\,K_{ij}\right)+4\lambda''(D_i\phi) D_j\phi\,,
\end{eqnarray}
\end{subequations}
where $\Pi$ is the conjugate momentum of $\phi$, $N_i$ is the GR momentum constraint, $\Omega=\gamma^{ij}\Omega_{ij}$, and $\Omega_i$ and $\Omega_{ij}$ come from the $3+1$ decomposition of the Weyl tensor.

The contributions from the kinetic and potential terms are given by the usual minimally coupled, real scalar terms,
\begin{eqnarray}
\rho^{\text{SF}} &=& \frac{1}{2}\Pi^2 + V(\phi) + \frac{1}{2}(D_i\phi)(D^i\phi) ~,\\
S_i^{\text{SF}} &=& -\Pi\, D_i \phi ~.
\end{eqnarray}

As mentioned in the introduction, previous work in such theories has required that $\rho^{\text{GB}} + \rho^X \ll \rho^{\text{SF}}$ and $S_i^{\text{GB}} + S_i^X \ll S_i^{\text{SF}}$ everywhere. This reduces the problem to (approximately) satisfying the regular GR constraint equations, and still allows the solution to evolve away from that of GR during the evolution. However, here we are able to treat the deviations from GR fully non-perturbatively.

\subsection{Choice of components to solve for}

As discussed in the introduction (and in more detail in \cite{Aurrekoetxea22}), in contrast to the CTT method that imposes a spatially constant value of the mean curvature $K$, and solves for the conformal factor $\psi = (\det{\gamma})^{1/12}$, the CTTK method allows both $K$ and $\psi$ to have a spatially varying profile. The momentum constraint (now with an additional source term from the spatial variation of $K$) is solved as in the CTT method, for the vector potential $W_i$, which is further decomposed into a scalar $U$ and a vector $V_i$. This additional flexibility in $K$ can be used to absorb some of the problematic matter source terms, resulting in a more robust method for general field configurations. The same approach will be used here, but we will see that in general the $4\partial ST$ terms are better absorbed into the elliptic equation for $\psi$, rather than in the mean curvature $K$.

The transverse-traceless part of the extrinsic curvature, $\bar{A}^{TT}_{ij}$, is usually set to zero, which roughly corresponds to a spacetime containing no gravitational waves. The conformal metric is also assumed to be $\delta_{ij}$ for simplicity (i.e., conformal flatness is assumed). It should be possible to apply the same techniques with more general choices of $\bar{A}^{TT}_{ij}$ and $\bar\gamma_{ij}$, such as those that are required for highly spinning black hole initial data, which would simply result in additional source terms in the equations. For now we maintain these choices for simplicity.

\subsection{Black hole spacetimes}\label{sec:black_hole_method}

With the $4\partial ST$ terms included and an appropriate form chosen for $U$ (see the appendix of \cite{Aurrekoetxea22} for a discussion), we write the Hamiltonian and momentum constraints in CTTK as follows
\begin{align}
    K^2&=24\pi\rho^{\text{SF}} \label{eq:K_squared} ~,\\
    \partial_j \partial^j \psi &= -\frac{1}{8}\psi^{-7}\Bar{A}^{ij}\Bar{A}_{ij}-2\pi \psi^{5} (\rho^{\text{GB}}+\rho^X)\label{eq:psi_elliptic} ~,\\
    \partial_j \partial^j V_i &= \frac{2}{3}\psi^6 \partial_i K + 8\pi \psi^6 (S^{\text{SF}}_i+S^{\text{GB}}_i+S_i^X) ~.\label{eq:Vi_elliptic} 
\end{align}
The GR case with $\rho=\rho^{\text{SF}}$ and $S_i = S_i^{\text{SF}}$ is derived in \cite{Aurrekoetxea22} - in what follows we explain the motivation for the positioning of the additional terms.

In regions of physical relevance, we can demand from an effective field theory (EFT) perspective that $\rho^{\text{GB}}+\rho^X < \rho^{\text{SF}}$. However, in some regions of high curvature this may not hold. Moreover, the value of $\rho^{\text{GB}}+\rho^X$ is not positive definite, and in practise for most cases of interest it varies between positive and negative regions in different parts of the spatial slice.
For this reason, $\rho^{\text{SF}}$ and $\rho^{\text{GB}}+\rho^X$ have been separated between equations \eqref{eq:K_squared} and \eqref{eq:psi_elliptic}. 
By including only the positive definite part in Eq. \eqref{eq:K_squared}, we avoid the possibility of $K^2$ having a negative source, even in regions where the additional curvature contributions dominate. However, since it is also not negative definite, the $\rho^{\text{GB}}+\rho^X$ term in equation \eqref{eq:psi_elliptic} is likely to appear with the ``wrong sign'' in a linearised expansion of $\psi$ in certain regions, which violates the conditions of the maximum principle and removes the guarantee of unique solutions \cite{Baumgarte06}. In all our test cases (see Section \ref{sec:tests}), this has not caused issues, and no further modification has been necessary for convergence to a solution. We speculate that this may be because we also include the contribution from $\Bar{A}^{ij}\Bar{A}_{ij}$ in the elliptic equation for $\psi$, which tends to stabilise the solutions
\footnote{This potential problem could be avoided by amending the split further such that $\rho^{\text{SF}} \rightarrow \rho^+ = \rho^{\text{SF}} + \rho^{P}$ and $\rho^{\text{GB}}+\rho^X \rightarrow \rho^- = \rho^{\text{GB}}+\rho^X - \rho^{P}$, with any $\rho^{P}$ that satisfies $\rho^{P} > 0$ and $\rho^{P} > \rho^{\text{GB}} + \rho^X$ everywhere, and $\rho^{P} \rightarrow 0$ at the boundaries.}.

If the spacetime contains one or more black holes, divergences will appear in $\psi$ and $\bar A_{ij}$. This can be dealt with by decomposing $\bar A_{ij}$ (and therefore $U$ and $V_i$) into a black hole part, $\bar A^{bh}_{ij}$, which contains the divergences, and a regular part, $\bar A^{reg}_{ij}$, which is solved for. The conformal factor $\psi$ can also be decomposed into a sum of isolated black-hole solutions, $\psi_{bh}$, and the regularised remainder, $\psi_{reg}$. 
The Poisson equations for $\psi$ and $V_i$ can then be solved directly, or by linearising around the previous solution and solving for the error. For example, the results given in Section \ref{sec:black_hole_tests} are calculated by fully solving for $V_i$ at each step, and iterating for $\psi$ as $\psi = \psi_0 + \delta\psi$.

The forms of $A^{bh}_{ij}$ and $\psi_{bh}$ are given in \cite{Aurrekoetxea22}, and are those proposed by Bowen and York in \cite{Bowen79,BowenYork80}.

\subsection{Cosmological spacetimes}

For simulations of cosmological spacetimes, periodic boundary conditions are often used (see \cite{East:2015ggf,Corman:2022alv,CloughInflation1, CloughInflation2, CloughInflation3,Bentivegna:2013xna,Widdicombe:2018oeo,Yoo:2018pda,Cook:2020oaj,Ijjas:2020dws,deJong:2021bbo}, although other approaches are possible \cite{Corman:2022rqo}). In the CTTK method without a $4\partial ST$ term, the method for solving the constraints with periodic boundaries is very similar to that for black hole spacetimes. Again, the constraints reduce to Equations \eqref{eq:K_squared} to \eqref{eq:Vi_elliptic}. However, the elliptic equations now impose a set of integrability conditions, as discussed in \cite{Garfinkle:2020iup,Bentivegna:2013xna}.  In cosmological spacetimes, the differential operator in equation \eqref{eq:Vi_elliptic} has a non-trivial kernel. Therefore, by the Fredholm alternative, solutions to this equation (which are necessarily non-unique) will only exist if the source is in the orthogonal complement of the kernel -- i.e., in the adjoint. Since the kernel includes constants, it is necessary (but not sufficient) for the right hand side of the Poisson equation to equal zero when integrated over the entire grid. A similar requirement applies to the right hand side of equation \eqref{eq:psi_elliptic} when it is treated as a constant source. With no $4\partial ST$ term, this simply corresponds to requiring a periodic distribution of the scalar field, and ensuring that there is no net momentum flux through the box in any direction.
However, with $\lambda(\phi) \neq 0$, this is no longer sufficient, as there is no guarantee of the source terms including $\rho^{\text{GB}}$, $\rho^X$, $S^{\text{GB}}_i$ and $S^X_i$ averaging to zero across the grid, and no obvious way of choosing them such that this is always the case. 

In simple cases where only the elliptic equation for $\psi$ is problematic (e.g., with a simple sinusoidal profile for $\phi$ and no conjugate momentum $\Pi$), this can be solved by further dividing $K$ into two parts, one constant and one spatially varying, which we call respectively $K_{C}$ and $K_{GR}$. The `GR' part of $K$ satisfies equation \eqref{eq:K_squared}, as in the GR case, sourced by $\rho^{\text{SF}}$, that is
\begin{align}
    K_{GR}^2&=24\pi\rho^{\text{SF}} \label{eq:K_squared_cosmo} ~.
\end{align}
The constant part of $K$, $K_{C}$, can then be used to compensate for any violation of the integrability condition for $\psi$. This is achieved by setting $K_{C}$ to a value satisfying the condition
\begin{multline}\label{eq:KGB_quadratic}
    K_{C}^2\int \frac{\psi^5}{12}d\Omega + K_{C}\int\frac{\psi^5K_{GR}}{6}d\Omega \\- \int \Big( \partial_j \partial^j \psi + \frac{1}{8}\psi^{-7} \bar{A}_{ij}\bar{A}^{ij} + 2\pi\psi^5(\rho^{\text{GB}}+\rho^X)\Big) d\Omega = 0 ~.
\end{multline}
As a result of these choices, the equation for $\psi$ becomes
\begin{multline}\label{eq:psi_final}
    \partial_j \partial^j \psi = -\frac{1}{8}\psi^{-7}\bar{A}_{ij}\bar{A}^{ij} - 2\pi\psi^5(\rho^{\text{GB}}+\rho^X) + \frac{1}{12}\psi^5 K_{C}^2 \\+ \frac{1}{6}\psi^5K_{C}K_{GR}\,.
\end{multline}
Equation \eqref{eq:KGB_quadratic} will always have real solutions, unless $\rho^{\text{GB}}$ dominates over the combined contributions of $\rho^{GR}$ and $\bar{A}_{ij}\bar{A}^{ij}$ when averaged across the grid
\footnote{If this is true the weak-coupling condition will not be satisfied, so the solutions can probably not be stably evolved anyway. However, we note that real solutions for the initial data will still be available if the sign of $\lambda(\phi)$ is chosen to make the discriminant positive - although this adds a physical restriction on the coupling.}.

The integrability condition for the momentum constraint can also be spoiled by the presence of various nonlinear terms in $S_i^{\text{GB}}$. With a shift-symmetric or quadratic coupling and $\Pi=0$, many of these terms are simplified. A sinusoidal profile for $\phi$ then gives a contribution to $S_i^{\text{GB}}$ that satisfies this constraint. For a more general $S_i^{\text{GB}}$, removing the assumption of conformal flatness would provide another source in the constraints, potentially allowing for this contribution to be cancelled out by a judicious choice. In our tests below we restrict ourselves to showing that the method works in the simpler case, and leave more general conditions to future work.

Even once the source
in equation \eqref{eq:Vi_elliptic} is in the adjoint and the right hand side of \eqref{eq:psi_final} integrates to zero, the solution to the elliptic equations at each step suffers from non uniqueness -- the equations have multiple solutions, where solutions differ by the addition of a constant or linear term in the equation for the conformal factor \footnote{This lack of uniqueness arises when the right hand side of equation \eqref{eq:psi_final} is treated as a constant source, which happens at each non-linear iteration if we do not solve perturbatively for $\psi$.}, and one or more Killing vectors of the conformally flat metric in the case of $V_i$ (see \cite{Garfinkle:2020iup}). This is addressed by solving for their values perturbatively, starting with an initial guess and solving for a small correction $(\delta\psi,~\delta V^i)$ at each step. This perturbative treatment naturally generates a linear term in the equation for $\delta\psi$ that prevents the constant and linear modes from growing. The freedom in $\delta V^i$ can also be eliminated by adding a small linear coefficient to the Poisson equation for $\delta V^i$. 
The addition of the conformal Killing vectors in $V_i$ is unimportant as they do not change the resulting value of the extrinsic curvature $K_{ij}$ \cite{Garfinkle:2020iup}. However, any non uniqueness in $\psi$ has a physical consequence -- its value at the start of the iteration picks out the final uniquely chosen solution, and this in turn determines physical properties of the field -- e.g., the density of the gradient terms measured by normal observers.

\section{Tests}\label{sec:tests}

\begin{figure*}[t]
    \includegraphics[width=0.9\textwidth]{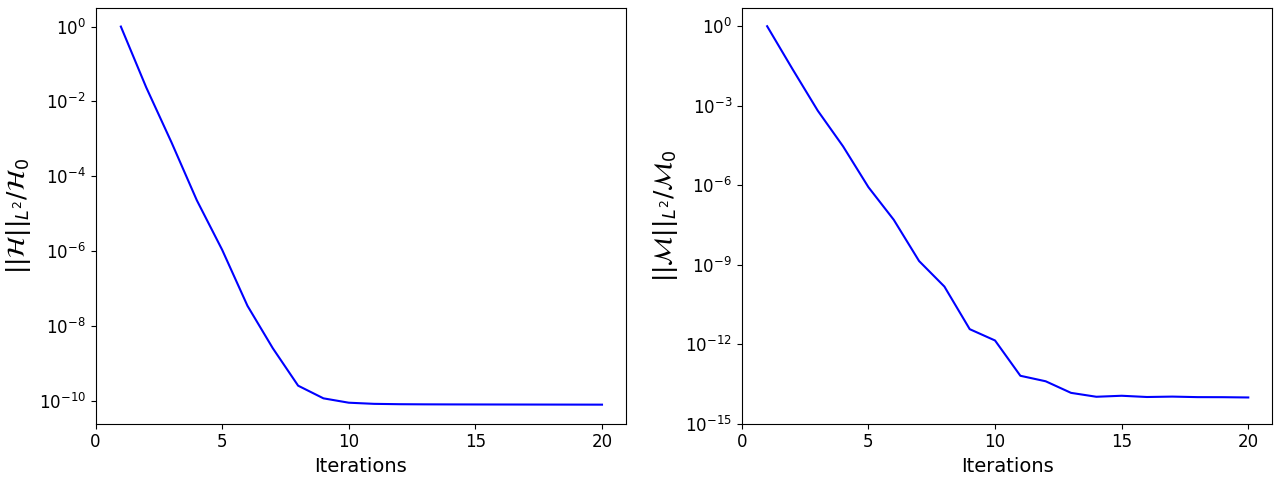}
     \caption{Plots of the $L^2$ norm across the grid of the Hamiltonian constraint violation ${\mathcal H}$ and the momentum constraint violation ${\mathcal M}=\sqrt{{\mathcal M}_i{\mathcal M}^i}$, normalised to their initial values ${\mathcal H}_0$ and ${\mathcal M}_0$, against non-linear iterations of the elliptic solver for the black hole spacetime with a dumbbell scalar field configuration shown in Fig. 2(b). We see that the solver converges to a good solution within ten iterations. The behavior is similar for the cosmological initial data.}
     \label{fig:iterations_convergence}
\end{figure*}

The methods described above for both black hole and cosmological spacetimes have been tested with a modified version of the CTTK solver used in \cite{Aurrekoetxea22}. This solver is constructed using the open-source Chombo \cite{Adams:2015kgr} code for finite difference solution of PDEs with Adaptive Mesh Refinement - in particular here we adapt their multigrid solver for elliptic equations. The results are imported into GRFolres \cite{joss_modified}, the modified version of the NR evolution code GRChombo \cite{Clough:2015sqa,Andrade:2021rbd}, to check the constraint violation using the methods verified in \cite{AresteSalo2022,AresteSalo23}. 

Here we show typical results for both overall convergence to a solution and convergence to the true zero-constraints solution with increasing resolution. The Chombo multigrid solver is designed to be second-order in all derivatives, so with the assumption that the errors are dominated by errors in the derivatives, the solver errors should also converge at that rate as the resolution is increased. This means that doubling the resolution should reduce the constraints by a factor of four
\footnote{Only two resolutions are required for the convergence tests since the true solution is known - i.e., the constraints should be zero across the grid.}. These tests have been conducted with a variety of coupling functions and potentials, although (as described above) the possible scalar field configurations are restricted in the case of periodic boundaries. 
Figure \ref{fig:iterations_convergence} shows the convergence of the constraint violations to zero (with a fixed resolution) as the number of non-linear iterations increases for the black hole spacetime described in Section \ref{sec:black_hole_tests}. This shows good convergence to a solution of the full non-linear problem within $\sim 10$ iterations.
In practise, this global measure is rather crude and ignores the fact that in some regions with small errors (often nearer the boundaries in BH spacetimes) the solver takes longer to show good local convergence. More information can be gained by checking the spatial profiles.
The convergence tests with increasing resolution for two particular cases are shown below in Figures \ref{fig:plots_bh} and \ref{fig:plots_periodic}. These plots show that the solver is consistently displaying the desired second-order convergence, but in the black hole case the solver was run for approximately 1000 non-linear iterations. Given that the solver is much less costly to run than the evolution code, such a high number of steps is not prohibitive -- with three levels of refinement, this takes a few hours with $\sim100$ CPU cores on a typical computing cluster. In the cosmological case, where only one level is used, it takes under an hour.

\begin{figure*}[t]
     \centering
     \begin{subfigure}{0.49\textwidth}
         \centering
        \includegraphics[width=1.0\textwidth]{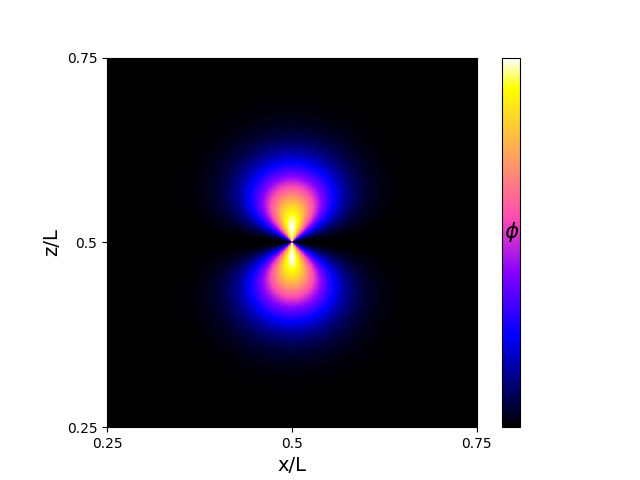}
         \caption{Dumbbell scalar field configuration around central black hole - this profile is used for both the field and its conjugate momentum.}
         \label{fig:phi_dumbbell}
     \end{subfigure}
     \hfill
     \begin{subfigure}{0.49\textwidth}
         \centering
         \includegraphics[width=1.0\textwidth]{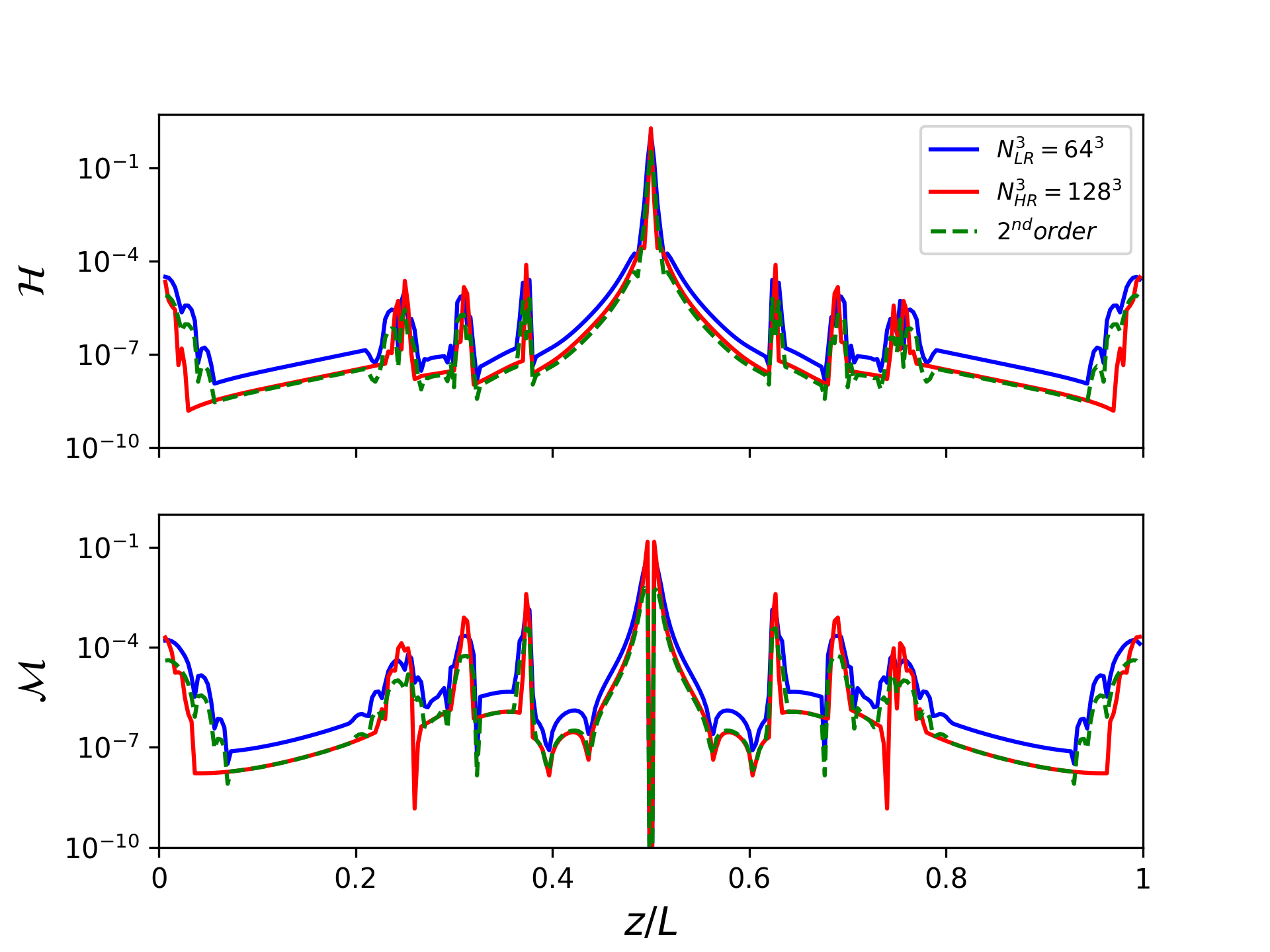}
         \caption{Local values of the Hamiltonian and momentum constraint violations along a line for two different resolutions, showing second order convergence.}
         \label{fig:convergence_dumbbell}
     \end{subfigure}
        \caption{Scalar field configuration and convergence plots for a growing dumbbell scalar field configuration around a spinning black hole.}
        \label{fig:plots_bh}
     \centering
     \begin{subfigure}{0.49\textwidth}
         \centering
         \includegraphics[width=1.0\textwidth]{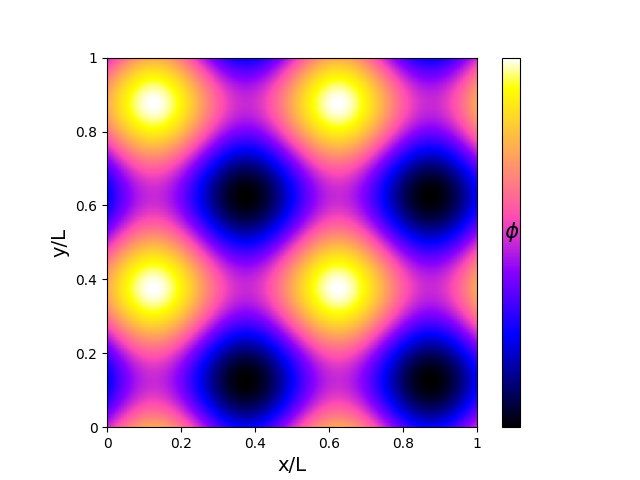}
         \caption{Sinusoidal scalar field configuration with periodic boundaries.  Here the conjugate momentum of the field is set to zero.}
         \label{fig:phi_periodic}
     \end{subfigure}
     \hfill
     \begin{subfigure}{0.49\textwidth}
         \centering
         \includegraphics[width=1.0\textwidth]{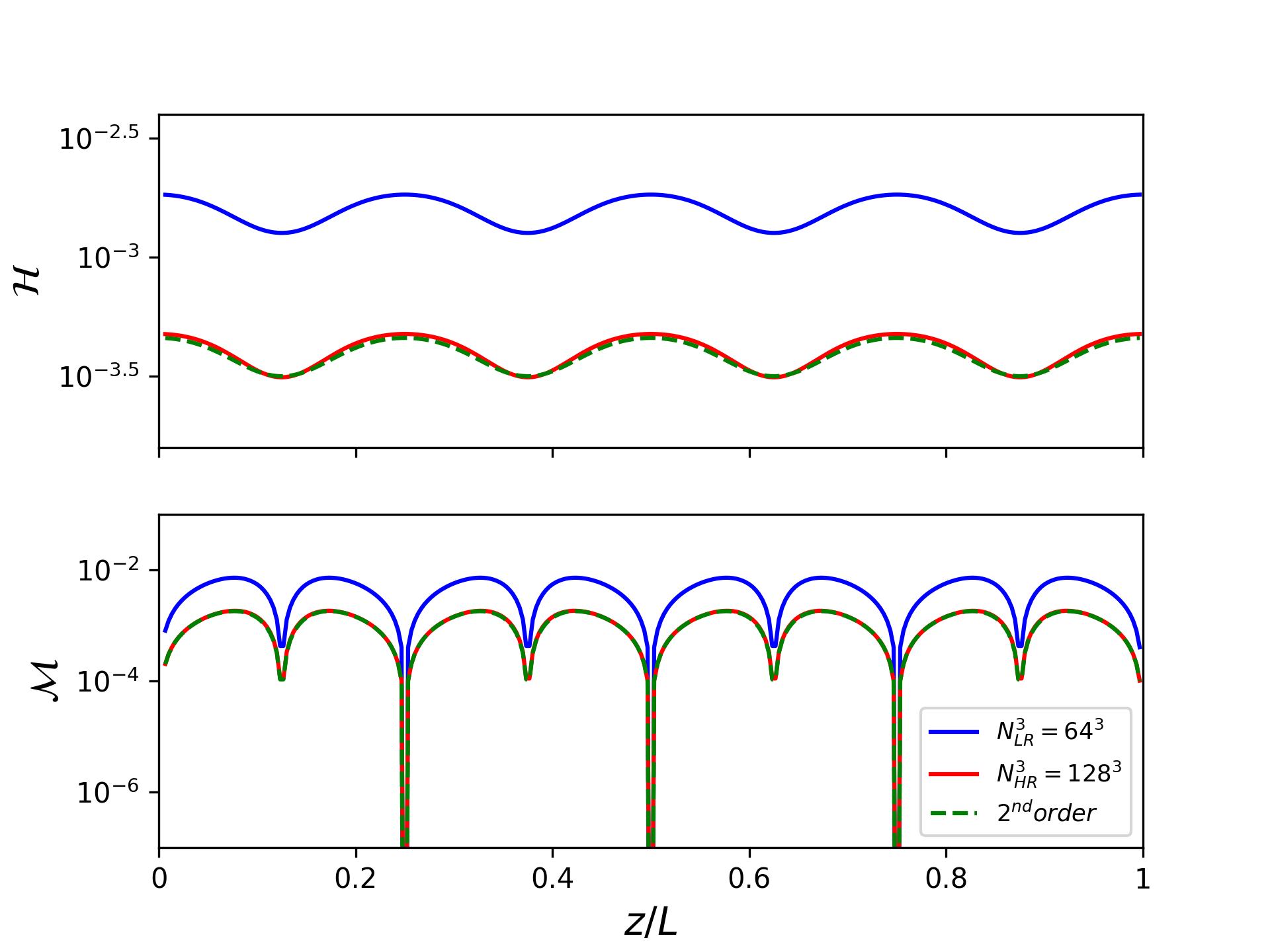}
         \caption{Local values of the Hamiltonian and momentum constraint violations along a line for two different resolutions, showing second order convergence.}
         \label{fig:convergence_periodic}
     \end{subfigure}
        \caption{Scalar field configuration and convergence plots for a sinusoidal scalar field configuration and periodic boundary conditions.}
        \label{fig:plots_periodic}
\end{figure*}

\subsection{Tests of black hole method}\label{sec:black_hole_tests}

The black hole method described in Section \ref{sec:black_hole_method} has been tested with a dumbbell-shaped scalar field and momentum configuration around a central black hole with dimensionless spin coefficient $a/M=0.5$ and a spin-axis along the $z$ direction, where $M$ is the total bare mass.
The potential and coupling functions are both chosen to be quadratic, i.e., $\lambda(\phi)=\frac{1}{2}\lambda_{\text{GB}}^2\phi^2$ and $V(\phi)=\frac{1}{2}m^2\phi^2$, with $m=1$ and $\lambda_{\text{GB}}/M=1$, a scalar field amplitude of $0.1$, and a momentum amplitude of $0.01$. In this test $g_2(\phi)$ was set to zero - its impact on the solutions was found to be negligible. This configuration was inspired by the scalar field configuration that the field has been found to settle into in spin induced cases --see for example the plots in \cite{AresteSalo23, Doneva23}-- although in such cases the field is massless. We do not attempt to match the stationary configuration exactly since this is simply a proof of principle, and we have tested other spatial configurations that show similar results to the ones here. Figure \ref{fig:phi_dumbbell} shows the amplitude of the scalar field across a slice of constant $y$-coordinate, passing through the singularity. Figure \ref{fig:convergence_dumbbell} then shows the convergence as the grid spacing is halved. The results with a finer grid are approximately second-order across the grid, other than at grid boundaries. In the Hamiltonian constraint the errors are not fully dominated by the derivatives, which is likely to cause the small deviation from exact second-order convergence in the central region. The convergence towards a solution for a fixed resolution is also shown, in Figure \ref{fig:iterations_convergence}.

\begin{figure}[h]
    \includegraphics[width=0.49\textwidth]{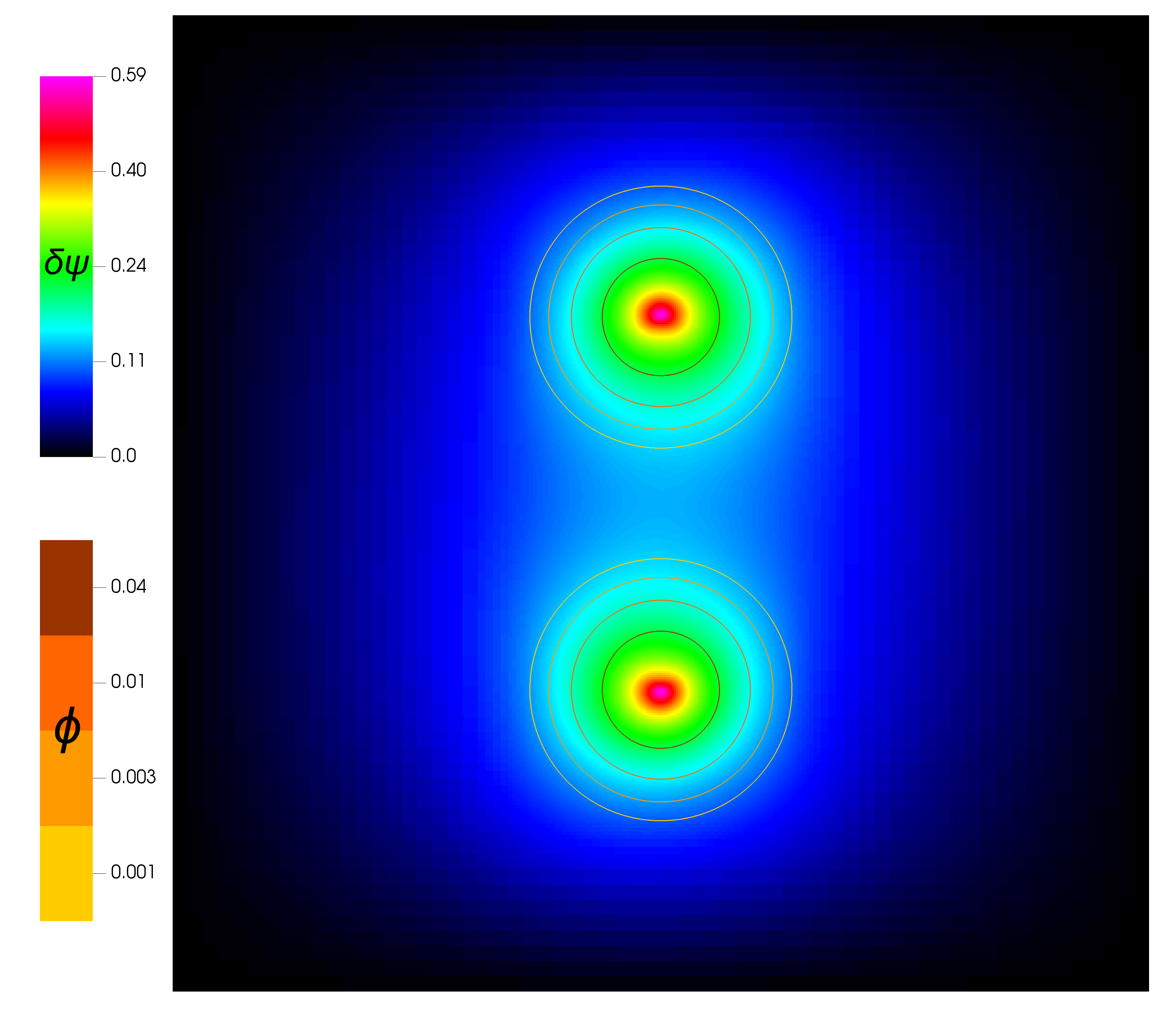}
     \caption{Correction to the conformal factor of the bare punctures for a black hole binary, with a superimposed contour plot showing the Gaussian scalar field over each black hole.}
     \label{fig:bh_binary}
\end{figure}

This method has also been tested with initial data for a black hole binary. The same coupling functions and amplitudes as the dumbbell test were used for two black holes with Bowen-York masses $m_{1,2}=0.5M$, momenta $|P|=2M$ perpendicular to their separation of $12M$, and dimensionless spin parameters $a/M=0.6$. In this case we set $g_2(\phi)/M^2 = 1$, although as above this only has a small impact on the solutions. A Gaussian scalar field with momentum was imposed over each black hole, as shown by the contours in Figure \ref{fig:bh_binary}, along with the change in the conformal factor over that of the bare punctures.

\subsection{Tests of cosmological spacetimes method}\label{sec:tests_cosmo}
A similar test was conducted with periodic boundary conditions, and a sinusoidal profile for $\phi$ in each direction. The same coupling and potential functions were used, with $m=0.5$ and the other parameters unchanged. As explained above, we needed to set the momentum of the field $\Pi$ to be zero in this case. Figure \ref{fig:plots_periodic} shows the scalar field configuration and tests of convergence with increasing resolution in the periodic case, and also displays second-order convergence across the grid. The speed of convergence with non-linear iterations is similar to that in the black hole case.

\section{Discussion and future work}

In this work the CTTK method was adapted to solve the constraint equations of $4\partial ST$ gravity, treating the additional curvature terms as another source to the Poisson equations of the GR case, as suggested in \cite{Kovacs:2021lgk}. Whilst such a treatment is a sensible first guess given that the new terms should be small in an effective theory (and given that in the weakly coupled regime a unique solution has been shown to exist \cite{Kovacs:2021lgk}), it is far from clear that such an approach will work in practise given the high non-linearity of the problem, especially as the coupling is made large. We have demonstrated that this is the case, and that the method is robust, provided certain choices are made about the split of the new terms between the available degrees of freedom. In fact, the method appears to be robust up to the strongly coupled regime of the theory, with coupling parameters of order 1. Beyond this regime the well-posedness of the evolution is no longer guaranteed, and in practise it will break down \cite{East:2020hgw,East:2021bqk,East:2022rqi,R:2022hlf, Doneva23,Thaalba:2023fmq}; therefore the solutions beyond this point are not of particular physical interest.

This adapted method has been tested for both black hole and cosmological spacetimes, and shows appropriate convergence in both cases. Techniques have also been described for guaranteeing the existence and uniqueness of solutions with either asymptotically flat or periodic boundary conditions.

This is the first method that has been demonstrated to work for fully satisfying the constraint equations for generic initial data in 3+1 dimensions, and it therefore expands the possibilities for numerical investigation of these theories. Future applications of this technique may include simulations of black holes with non-trivial initial scalar field configurations, and tests of inflation in $4\partial ST$ gravity with inhomogeneous initial conditions. It would also be useful to extend the method to permit non conformally flat spacetimes (which could address the issues with integrability of the momentum constraints that we encountered in the cosmological case) and to check that a similar approach works in the alternative extended conformal thin sandwich (XCTS) method \cite{York:1998hy,Pfeiffer:2002iy}, in which one solves also for the lapse and the shift functions to achieve a specific initial time evolution of the variables. XCTS is more widely used in practise than CTT and offers a number of advantages, especially for identifying equilibrium initial data. We anticipate that the approach developed here should work equally well in such cases, but we hope to see this demonstrated in future work by groups working with such solvers.

\section{\label{sec:acknowledge}Acknowledgements}

We would like to thank Josu Aurrekoetxea and Eugene Lim for helpful discussions about the CTTK method, and for the use of the CTTK code they developed with KC, which was modified in this work. We also thank Aron Kov\'acs and Miguel Bezares for helpful conversations.
We thank the entire \texttt{GRChombo} \footnote{\texttt{www.grchombo.org}} collaboration for their support and code development work. PF would like to thank the Enrico Fermi Institute and the Department of Physics of the University of Chicago for hospitality during the final stages of this work. PF is supported by a Royal Society University Research Fellowship  No. URF\textbackslash R\textbackslash 201026, and No. RF\textbackslash ERE\textbackslash 210291. KC is supported by an STFC Ernest Rutherford fellowship, project reference ST/V003240/1. PF and KC are supported by an STFC Research Grant ST/X000931/1 (Astronomy at Queen Mary 2023-2026). SB is supported by a QMUL Principal studentship. LAS is supported by a QMUL Ph.D. scholarship. 
APS thanks Queen Mary University of London for hosting him to work on this project, and Erik Schnetter for discussions on initial condition solvers.
This work used the ARCHER2 UK National Supercomputing Service (https://www.archer2.ac.uk).
This work also used the DiRAC@Durham facility managed by the Institute for Computational Cosmology on behalf of the STFC DiRAC HPC Facility (www.dirac.ac.uk). The equipment was funded by BEIS capital funding via STFC capital grants ST/P002293/1, ST/R002371/1 and ST/S002502/1, Durham University and STFC operations grant ST/R000832/1. DiRAC is part of the National e-Infrastructure.
Calculations were also performed using the Sulis Tier 2 HPC platform hosted by the Scientific Computing Research Technology Platform at the University of Warwick. Sulis is funded by EPSRC Grant EP/T022108/1 and the HPC Midlands+ consortium. This research also utilised Queen Mary’s Apocrita HPC facility, supported
by QMUL Research-IT \cite{apocrita}.
For the purpose of Open Access, the author has applied a CC BY public copyright licence to any Author Accepted Manuscript version arising from this submission.

\bibliography{main}

\end{document}